# A Semi-Analytical Approach to Model Drilling Fluid Leakage Into Fractured Formation

Rami Albattat, Hussein Hoteit*

King Abdullah University of Science and Technology (KAUST)

## Abstract

Loss of circulation while drilling is a challenging problem that may interrupt operations, reduce efficiency, and contaminate the subsurface formation. When a drilled borehole intercepts conductive faults or fractures, lost circulation manifests as a partial or total escape of drilling, workover, or cementing fluids, into the surrounding rock formations. Loss control materials (LCM) are often used in the mitigation process. Understanding the fracture effective hydraulic properties and fluid leakage behavior is crucial to mitigate this problem. Analytical modeling of fluid flow in fractures is a tool that can be quickly deployed to assess lost circulation and perform diagnostics, including leakage rate decline, effective fracture conductivity, and selection of the optimum LCM. Such models should be applicable to Newtonian, as well as non-Newtonian yield-stress fluids, where the fluid rheology is a nonlinear function of fluid flow and shear stress. In this work, a new semi-analytical solution is developed to model the flow of non-Newtonian drilling fluid in a fractured medium. The solution model is applicable for various fluid types exhibiting yield-power-law (Herschel-Bulkley). We use high-resolution finite-element simulations based on the Cauchy equation to verify our solutions. We also generate type-curves and compare them to others in the literature. We then demonstrate the applicability of the proposed model for two field cases encountering lost circulations. To address the subsurface uncertainty, we combine the semi-analytical solutions with Monte-Carlo simulations and generate probabilistic predictions. The solution method can estimate the range of fracture conductivity, parametrized by the fracture hydraulic aperture, and time-dependent fluid loss rate that can predict the cumulative volume of lost fluid. The proposed approach is accurate and efficient enough to support decision-making for real-time drilling operations.

## Introduction

Naturally fractured formation is often prone to severe mud loss during drilling operations. As a result, problems may emanate because of the severity of lost fluid such as kicks, wellbore instability, formation damage, and sometimes freshwater aquifer contamination (Seyedmohammadi 2017). One procedure to encounter and mitigate this problem is to add lost circulation material (LCM) to the drilling fluid. LCM is commonly used in drilling applications to reduce and stop lost circulation (Attong, Singh, and Teixeira 1995; Ali, Kalloo, and Singh 1997; Olsen et al. 2019; Knudsen et al. 2015). The fluid characteristics of LCM, such as density and viscosity should be carefully selected based on the formation hydraulic properties such as the conductivity of the thief layers and fractures, among other factors (Luzardo et al. 2015). Due to the time-scale of the problem, there is a need to develop accurate and efficient modeling tools applicable to real-time drilling operations to perform diagnostics and predictions. Field observations suggest that, during mud

hussein.hoteit@kaust.edu.sa

filtration into porous media, an immediate spurt loss generally occurs, which then decreases in rate as filter cake is being deposited.

On the other hand, mud loss rate into fractures often exhibits a sudden peak, followed by a gradual declining loss (Dyke, Bailin Wu, and Milton-Taylor 1995). The transient rate decline is related to the fluid pressure build-up within the fracture. The ultimate lost volume is a function of the fluid mobilities, and the fracture conductivity, pore-volume, and extension (NORMAN 2011). Analytical solutions for mud loss into a single effective fracture, mimicking a fractured formation has been studied for decades in the literature. Early modeling attempts for simplified cases were based on Darcy's Law at steady-state conditions (Bannister and Lawson 1985; Bruckdorfer and Gleit 1988). Sanfillippo et al. (1997) introduced a semi-analytical solution for Newtonian fluid flow into a horizontal fracture by combining the diffusivity equation and mass conservation in one-dimensional (1D) radial systems. The derived ordinary differential equation (ODE) was solved numerically. Maglione and Marsala (1997) presented an analytical solution of the diffusivity equation for fluids with a constant viscosity at the steady-state conditions. Liétard et al. (2002) developed type-curves based on numerical solutions to describe time-dependent mud-loss volumes into a horizontal fracture. The model is applicable for non-Newtonian fluids exhibiting Bingham-Plastic rheological behavior. The authors generated type-curves, based on dimensionless groups, to describe mud-loss rates as a function of the fracture hydraulic aperture and fluid properties. Other authors (Liétard et al. 2002, 2002) proposed analytical approaches to generate similar type-curves to the ones proposed by Liétard et al. (2002). Huang et al. (2011) derived a method to estimate the hydraulic fracture aperture by simplifying insignificant terms in the governing equations. Majidi et al. (2010) generated approximate analytical solutions for yield-power-law fluids by reducing a Tylor expansion of the governing nonlinear flow equation into its linear terms. This approximation helped to generate an analytical solution but was found to introduce inaccuracies in some cases, as discussed in the paper. Motivated by the work of Majidi et al. (2010), Dokhani et al. (2020) introduced a mathematical model and numerical solutions to account for fluid leak-off from fractures. Other authors proposed various numerical methods based on higher-order discretizations (Ambartsumyan et al. 2019; Girault and Rivière 2009; Arbogast and Brunson 2007; Shao et al. 2016; Ţene, Al Kobaisi, and Hajibeygi 2016).

In this work, a new semi-analytical solution is developed to model the flow of non-Newtonian drilling or LCM fluids exhibiting a yield-power-law (Herschel-Bulkley) behavior. The non-Newtonian fluid flow is described by the Cauchy momentum equation. The nonlinear system of equations is reformulated and converted into a system of ODE's, which is then solved numerically with an efficient ODE solver (Hindmarsh et al. 2005). For convenience, we also introduce dimensionless groups and develop type-curves, which describe fluid volume loss behavior versus dimensionless time, as a function of the fracture and fluid properties. The fracture is represented by two parallel plates in a 1D radial system. We use high-resolution finite element simulations from commercial software, COMSOL (Littmarck and Saeidi 1986), to verify our model. We also compare our proposed semi-analytical solutions to two other models from the literature. The developed type-curves describe the mud-loss volume and mud-invasion front velocity under various fluid rheological properties, drilling pressure conditions, and fracture aperture. We use the type-curves and demonstrate the applicability of the proposed model for two field cases. We discuss a simple approach to combine the developed semi-analytical solutions with Monte-Carlo simulations to address uncertainties.

This paper is organized as follows; we first review the main governing equation for non-Newtonian fluid flow. Then, we reformulate the equations for 1D radial system, followed by a discussion of the solution method of the obtained semi-analytical system. In section two, we compare our solution to other analytical solutions for a particular fluid-type case. For general cases, we verify our solutions with full-physics numerical simulations. In the third section, we introduce new type-curves with the corresponding dimensionless groups. In section four, before the conclusion, we demonstrate the applicability of the proposed approach for two field cases.

## Physical & Mathematical Model

The general governing equation used to describe non-Newtonian fluid dynamics is given by Cauchy equation (Irgens 2014; Cioranescu, Girault, and Rajagopal 2016), such that,

$$\rho \frac{\partial \mathbf{v}}{\partial t} + \rho (\mathbf{v} \cdot \nabla) \mathbf{v} = \nabla \cdot (-p\mathbf{I} + \boldsymbol{\tau}) + \rho \mathbf{g} \tag{1}$$

In the above equation, the transient term consists of the fluid density $\rho$, velocity vector $\mathbf{v}$, and time $t$. The fluid pressure is denoted by $p$, the fluid shear stress by $\boldsymbol{\tau}$, the gravitational acceleration by $\mathbf{g}$, and the identity matrix by $\mathbf{I}$.

The fracture is represented by two parallel radial plates, perpendicular to the wellbore, as shown in **Fig. 1**. Note that horizontal fractures could occur at shallow depths and over-pressurized formations (Smith and Montgomery 2015; Ben-Avraham et al. 2012).

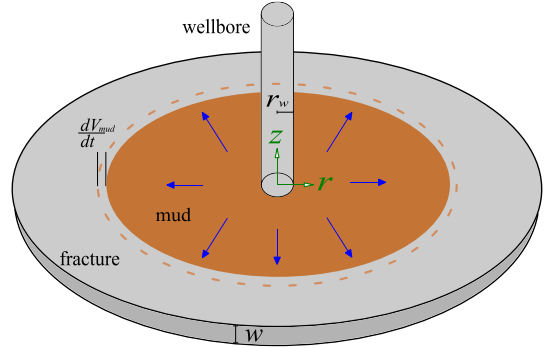

**Fig. 1**–Physical domain mimicking a horizontal fracture intercepting the wellbore. The shaded brown area shows the mud radial invasion, $r_w$ is wellbore radius, w is fracture aperture, $V_{mud}$ is mud-loss volume.

Assuming steady-state conditions and neglecting gravity with low inertial effect in comparison to other forces, Eq. (1) becomes,

$$0 = \nabla \cdot (-p\mathbf{I} + \boldsymbol{\tau}). \tag{2}$$

In 1D radial system, the above equation simplifies to (Panton 1984),

$$\tau(z, r) = z \frac{\partial p}{\partial r}. \tag{3}$$

Where $\tau(z, r)$ is the radial shear stress component, perpendicular to the $z$-direction.

On the other hand, the shear stress component $\tau$ is described by the Herschel-Bulkley fluid model (Hemphil, Pilehvari, and Campos 1993), that is,

$$\tau(z,r) = \tau_0 + m\left(\frac{dv_r}{dz}\right)^n. \qquad (4)$$

In Eq. (4), the parameter $\tau_0$ is the yield shear stress, which determines the fluidity state, as explained later. The flow index $n$ is a positive number that reflects fluid rheological behavior. For instance, the fluid exhibits shear-thinning behavior when $n<1$, and shear-thickening when $n>1$. Typical values for flow behavioral index in drilling fluid ranges from about 0.3-1.0 (Kelessidis et al. 2006). The other parameters correspond to the consistency multiplier $m$, and the derivative of the radial velocity $v_r$ in the z-direction, reflecting the shear rate. Note that the Herschel-Bulkley model in Eq. (4) can also be used to describe Newtonian fluids when considering $\tau_0 = 0$ and $n=1$.

## Solution Method

Our system of equations is given by the Cauchy equation (3) and the Herschel-Bulkley fluid model Eq.(4). The first equation combines two external forces, pressure force, and shear force, applied to the fluid. The second equation describes the fluid rheological behavior as a function of the two external forces. The proposed solution is based on the assumption that the drilling fluid viscosity is higher than the in-situ water viscosity in the fracture, with no significant mixing between the two fluids. Under these conditions, a piston-like displacement is considered at the mud-water interfaces, which is a reasonable assumption (Razavi et al. 2017). Therefore, the fluid pressure and the shear force are approximated only within the mud invaded zone. The fracture is assumed infinite acting with constant average hydraulic aperture, and no-slip boundary conditions (B.C.) are set at the fracture walls.

The concept of the proposed solution method is illustrated in **Fig. 2**. As the fluid propagates deep into the fracture, radial flow velocity decreases, and shear stress diminishes. Therefore, shear-thinning is highest in the vicinity of the wellbore and gradually reduces with radial distance, exhibiting a shear thickening behavior. Furthermore, driven by the vertical variations of flow velocity and shear stress, fluid layers along the fracture aperture develop and introduce self-friction with the highest intensity at the fracture wall and reduces linearly towards the fracture centerline, as illustrated in Fig. 2. Consequently, a region with diminished shear rate (i.e. $dv_r/dz = 0$) develops at the fracture center, corresponding to the yield shear stress zone $\tau_0$ (see Eq. (4)). This region is denoted by the plug flow region, while the rest is the free flow region. As the fluid propagates further within the fracture, the plug region expands toward the fracture walls, eventually reaching a total stall of the flow, as illustrated in **Fig. 3a**. This condition represents the ultimate steady-state, where the pressure drop between the wellbore and the mud front becomes too small to overcome the yield stress $\tau_0$ (see Fig. 3b).

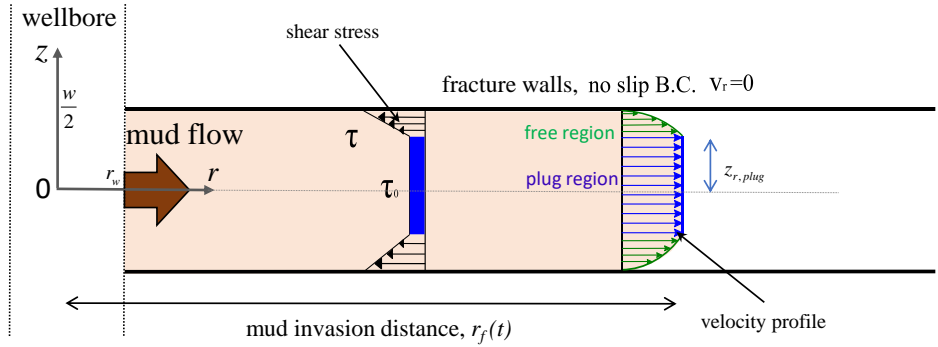

**Fig. 2**–Sketch of an infinite-acting fracture cross-section of aperture $w$, intercepting the wellbore. No-flow and no-slip B.C. are imposed at the fracture wall. $r_f(t)$ is the time-dependent radial distance of the invading fluid into the fracture; $v_r$ is the velocity vector, and $\tau$ is the shear stress. The plug flow region, resulting from shear-thickening, corresponds to the region where the variation of the velocity along the z-axis diminishes; the free flow region corresponds to the region outside the plug region.

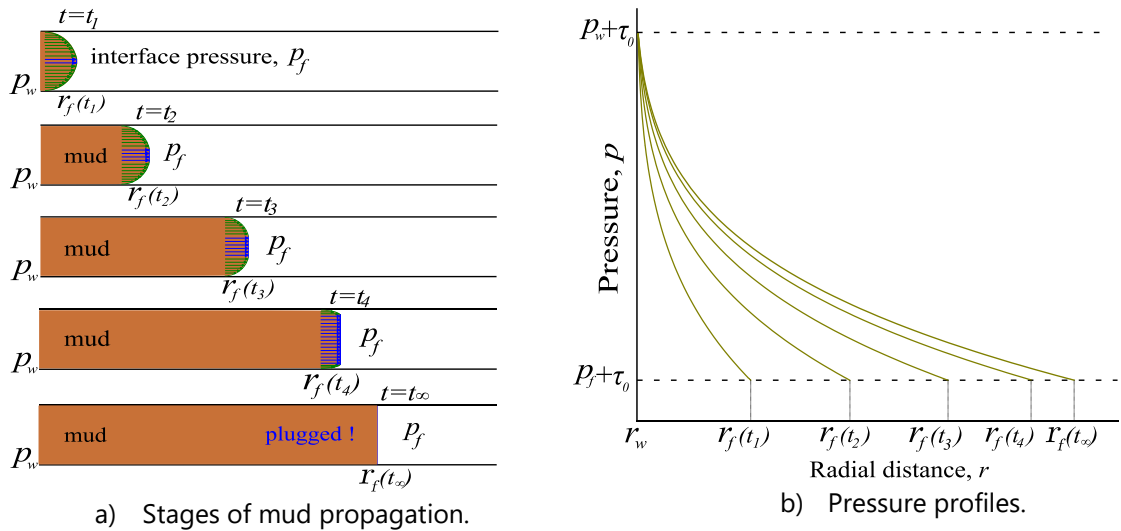

a) Stages of mud propagation.

b) Pressure profiles.

**Fig. 3**–Illustration (a) of yield-power-law fluid flow in a radial fracture showing the evolution of the propagation of the plug region as the mud travels away from the wellbore, resulting in total plugging. Plot (b) shows typical pressure profiles versus radial distance at various times and invasion distances.

The fluid flow in the fracture is symmetrical across the fracture centerline. Therefore, we derive the equations for the domain upper half of symmetry. Based on the previous discussion, where we subdivided the flow domain into two regions a plug and a free region, the following conditions are introduced for the velocity vector:

$$v_r(z) = \begin{cases} v_{r,plug}(z), & for\ z \leq z_{plug} \\ v_{r,free}(z), & for\ z_{plug} < z < \dfrac{w}{2} \\ 0, & for\ z = \dfrac{w}{2} \end{cases} \quad (5)$$

In Eq. (5), $z_{plug}$ is the vertical extension of the plug region. $v_{r,plug}$, and $v_{r,free}$ denote the velocities within the plug region and free region, respectively. The last condition is a result of no-slip B.C.

A detailed description of the solution method is provided in **Appendix A**. Here, we only show the key derivations. Combining Eqs. (3) and (4), and solving the differential equation of velocity in the z-direction, we get,

$$v_r(z) = \frac{n\left(-\dfrac{\partial p}{\partial r}\dfrac{w}{2}+\tau_0\right)\left(\dfrac{\dfrac{\partial p}{\partial r}\dfrac{w}{2}-\tau_0}{m}\right)^{1/n} + n\left(-\dfrac{\partial p}{\partial r}z+\tau_0\right)\left(\dfrac{\dfrac{\partial p}{\partial r}z-\tau_0}{m}\right)^{\frac{1}{n}}}{\dfrac{\partial p}{\partial r}(n+1)} \quad (6)$$

In the plug region, $dv_r/dz = 0$, therefore, writing Eq. (6) for each region separately, one gets,

$$v_{r,free}(z) = \frac{n}{n+1}\left(z_{plug}-\frac{w}{2}\right)\left(\frac{\dfrac{\partial p}{\partial r}\left(\dfrac{w}{2}-z_{plug}\right)}{m}\right)^{1/n} + \frac{n}{n+1}(z-z_{plug})\left(\frac{\dfrac{\partial p}{\partial r}(z-z_{plug})}{m}\right)^{\frac{1}{n}}$$

$$v_{r,plug}(z) = \frac{n}{n+1}\left(\frac{\tau_0}{\dfrac{\partial p}{\partial r}}-\frac{w}{2}\right)\left(\frac{\dfrac{w}{2}\dfrac{\partial p}{\partial r}-\tau_0}{m}\right)^{1/n} \quad (7)$$

The total volumetric flow rate $Q_{total}$, is written as the sum of the rate within the plug region $Q_{plug}$, and within the free region $Q_{free}$, that is,

$$Q_{total} = Q_{plug} + Q_{free} \quad (8)$$

On the other hand, the flux can be expressed in terms of the surface integral for each region by:

$$Q_{total} = 4\pi r \int_0^{z_{plug}} v_{r,plug}\,dz + 4\pi r \int_{z_{plug}}^{w/2} v_{r,free}\,dz \quad (9)$$

Substituting Eqs. (7) in Eq. (9) and integrating along the fracture aperture, we get,

$$Q_{total}^n = \frac{(4\pi r)^n}{m}\left(\frac{w}{2}\right)^{2n+1}\left(\frac{n}{2n+1}\right)^n\left(\frac{dp}{dr}\right)\left(1-\frac{\tau_0}{\frac{w}{2}\frac{dp}{dr}}\right)\left(1-\left(\frac{1}{n+1}\right)\frac{\tau_0}{\frac{w}{2}\frac{dp}{dr}}-\left(\frac{n}{n+1}\right)\left(\frac{\tau_0}{\frac{w}{2}\frac{dp}{dr}}\right)^2\right)^n \quad (10)$$

Be rearranging Eq. (10), it can be simplified to an ODE with a quadratic form, as follows,

$$\left(\frac{dp}{dr}\right)^2 - \left(\frac{Q_{total}^n}{r^n A}+B\right)\frac{dp}{dr}+D=0 \; , \quad (11)$$

where, the quantities $A, B,$ and $D$ are nonlinear functions of $n$, $m$, $w$, as discussed in **Appendix A**.

By solving for the quadratic equation (11) analytically and selecting the positive physical root, we get,

$$\frac{dp}{dr} = \frac{1}{2}\left(B+\frac{Q_{total}^n}{r^n A}+\sqrt{\left(B+\frac{Q_{total}^n}{r^n A}\right)^2-4D}\right) \quad (12)$$

Finally, by integrating Eq. (12) and recalling the total flux, we reach the final ODE system to solve:

$$\begin{cases} p_f - p_w = \dfrac{B\left(r_f(t)-r_w\right)}{2}+\dfrac{Q_{total}^n\left(r_f(t)^{1-n}-r_w^{1-n}\right)}{2(1-n)A}+\dfrac{1}{2}\int_{r_w}^{r_f(t)}\left(\sqrt{\left(B+\dfrac{Q_{total}^n}{r^n A}\right)^2-4D}\right)dr \\ Q_{total} = 2\pi w r_f(t)\dfrac{dr_f(t)}{dt} \end{cases} \quad (13)$$

The ODE system in Eq. (13) is nonlinear and cannot be solved analytically, except for a particular case when $n=1$ (see **Appendix B**). Therefore, a numerical ODE solver is used to solve Eq. (13) starting with initial conditions $r_f(t=0)=r_w$ (see **Appendix A**).

**Model Verification: Particular Analytical Solution.**
Our proposed semi-analytical solution is a generalization of the one introduced by Liétard et al. [(2002)] for Bingham plastic fluids. By taking a unity flow index ($n=1$) in Eq.(4), the Herschel-Bulkley model reduces to the Bingham plastic model such that,

$$\tau(z,r) = \tau_0 + m\frac{dv_r}{dz} \quad (14)$$

In **Appendix B,** we show the derivation of a particular case of the proposed solution when $n=1$. **Fig. 4** compares the solutions of the mud invasion front (dimensionless) versus a dimensionless time obtain by our model and the one proposed by Liétard et al. (2002), which shows identical match. The solutions are presented for various values between 0.002 and 0.04 of the dimensionless parameter $\alpha$, which is defined by,

$$\alpha = \frac{3r_w}{w}\left(\frac{\tau_0}{p_f - p_w}\right). \tag{15}$$

This test case shows that a particular solution ($n=1$) of the proposed semi-analytical model converges to the analytical solution of Liétard et al. (2002) for Bingham plastic fluids.

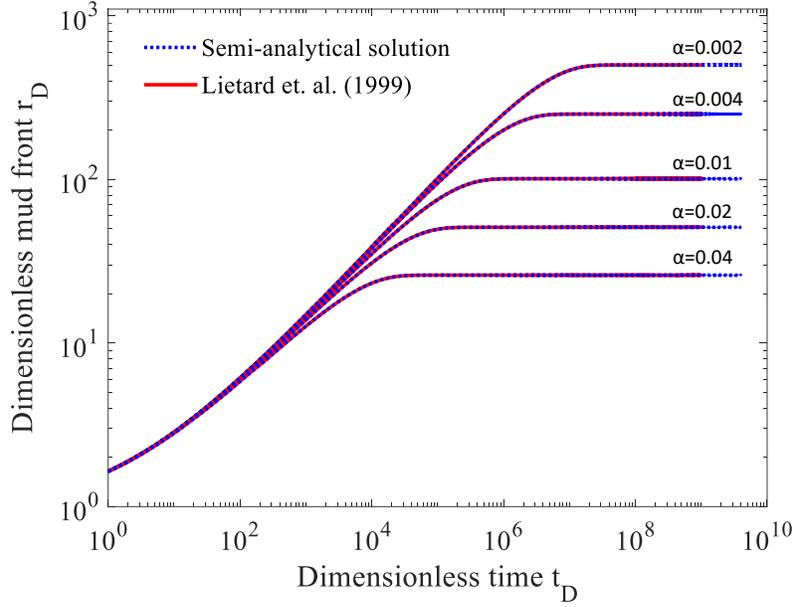

**Fig. 4**–Comparisons of solutions from a particular case of the proposed semi-analytical with the analytical solution by Liétard et al. (2002) for Bingham plastic fluid model. The type-curves corresponding show the invasion front versus time, generated for different values of alpha between 0.002 and 0.04.

**Model Verification: General Solution.**
We use simulations to verify our semi-analytical solutions for general Herschel-Bulkley fluids with different flow indexes, $n$. Simulations were performed using a finite-element method within COMSOL-Multiphysics (Littmarck and Saeidi 1986) to solve the flow of a non-Newtonian fluid within two parallel circular plates, mimicking the geometry of the radial fracture, shown in Fig. 1. We emphasize that the numerical solution is based on the Navier-Stokes equations, whereas the semi-analytical solution is based on the more general Cauchy equation of motion corresponding to the Herschel-Bulkley fluid. To address the discrepancy between the two solution methods, we use the approach by Papanastasiou [1987], which links the Navier-Stokes equation with the Cauchy equation by manipulating the Herschel-Bulkley fluid as,

$$\mu_{eff}(\gamma) = \mu_0\left(1 - e^{-m_p\gamma}\right) + m(\gamma)^{n-1} \tag{16}$$

Where, the effective viscosity as a function of shear rate, $\mu_{eff}(\gamma)$ is written in terms of the viscosity due to yield stress $\mu_0$, consistency multiplier $m$, shear rate $\gamma$, behavioral flow index $n$, and a regularization exponent $m_p$. The parameter $m_p$ has no physical significance. However, it is important to avoid singularities in the numerical solutions. Based on a sensitivity study (not shown here) to assess the effect of $m_p$, we found that $m_p$ =100 to 500 provided reasonable accuracy and robustness. High-resolution simulations were also needed to reduce instabilities and numerical artifacts. The accuracy of this numerical solution has been discussed previously (Albattat and Hoteit 2019).

In the simulation model, the fracture is assumed to be initially saturated in water at a constant pressure. The size of the fracture in the simulation domain is selected to be large enough to mimic infinite boundary conditions. The simulation solution includes the pressure solution in the mud and water zones, and the mud-front propagation (Albattat and Hoteit 2019). **Fig. 5** compares the semi-analytical solutions and the numerical solutions obtained for two cases with $n=1$, and $n=0.7$, reflecting a yield stress shear-thinning fluid. The solutions describe the mud invasion distance versus time. The plateau section in the curves reflect the maximum invasion distance, that is when the mud-front stalls. Both solution methods are in good agreement.

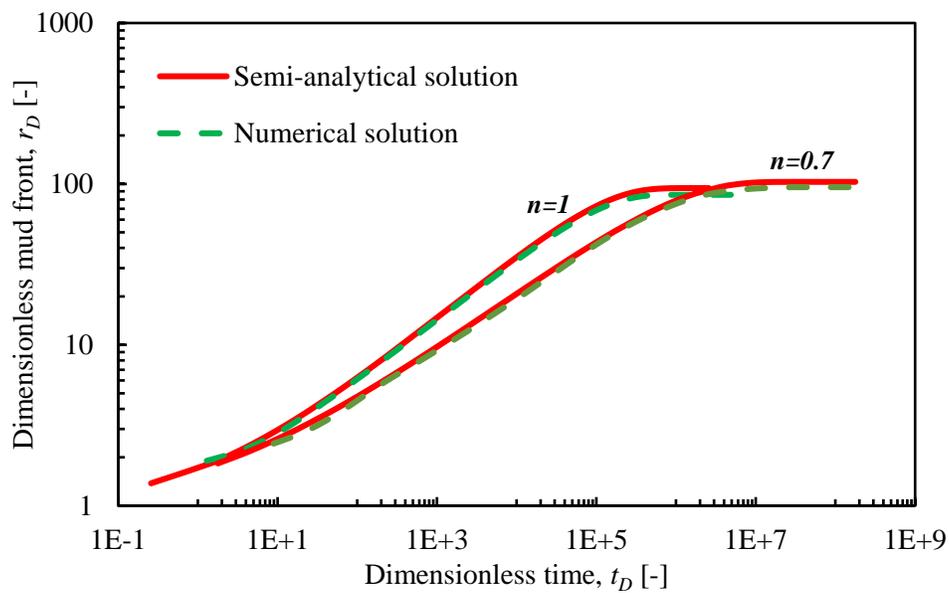

**Fig. 5**–Comparison between the proposed semi-analytical solutions with the numerical solutions by COMSOL. Both methods show good agreements for different cases, *n*=1, and 0.7.

## Type-Curves

Type-curves are often generated with dimensionless groups to provide quick interpretations and diagnostics of time-dependent trends. This technique is commonly used in well testing (Lee 1982). In this work, we adopt the following dimensionless variables:

$$r_D = \frac{r_f}{r_w}$$

$$V_D = \frac{V_m}{V_w} = \frac{\pi w(r_f^2 - r_w^2)}{\pi w r_w^2} = \left(\frac{r_f}{r_w}\right)^2 - 1 = r_D^2 - 1$$

$$\alpha = \left(\frac{2n+1}{n+1}\right)\left(\frac{2r_w}{w}\right)\left(\frac{\tau_0}{\Delta p}\right) \quad (17)$$

$$\beta = \left(\frac{n}{2n+1}\right)\left(\frac{w}{r_w}\right)^{1+\frac{1}{n}}\left(\frac{\Delta p}{m}\right)^{\frac{1}{n}}$$

$$t_D = t\beta$$

Where, $r_D$ is the dimensionless mud-invasion radius, $V_D$ is the dimensionless mud-loss volume, $t_D = \beta t$ is the dimensionless time, $\alpha$ is a dimensionless parameter reflecting the fluid rheological and invasion behavior. Additional analysis of the effect of $\alpha$ is discussed below. The governing system of equations given in Eq.(13) is then rewritten in terms of the dimensionless variables, where the same solution method is applied. Additional details on the derivation of the dimensionless form of the semi-analytical solution are shown in **Appendix C**. **Fig. 6** shows a set of type-curves generated for different flow indexes $n =$ 1.0, 0.8, 0.6, 0.4, and different values of $\alpha$. The type- curves show cumulative mud loss increase versus time. At far enough distance from the wellbore, the plugging effect becomes more significant, and the fluid eventually halts.

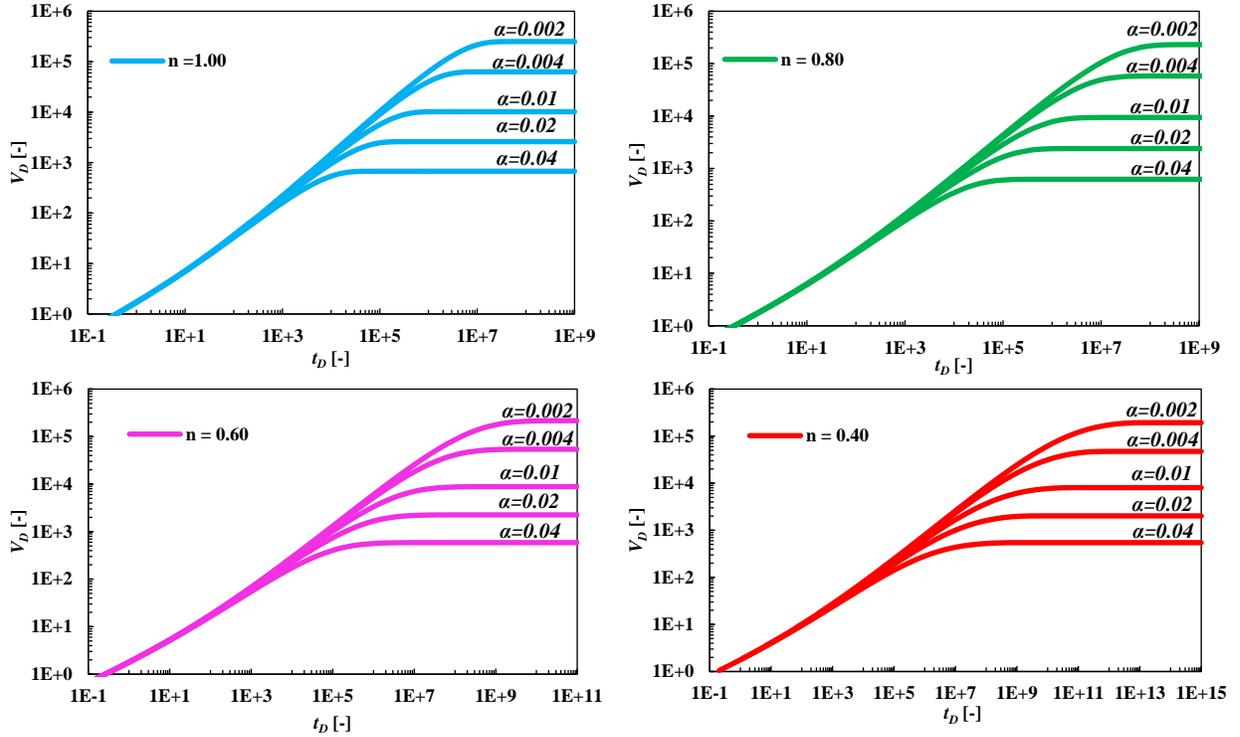

**Fig. 6**–Type-curves in a dimensionless form showing dimensionless mud-loss volume ($V_D$) versus dimensionless time ($t_D$) for different values of $n$ ($n$ =1, 0.8, 0.6, and 0.4) and $\alpha$ ($\alpha$ =0.002, 0.004, 0.01, 0.02, and 0.04).

We compare our derived type-curves with the one proposed by (Majidl et al. 2010). Majidi's et al. model simplifies a nonlinear term in the governing equation by a linear term corresponding to a first-order Taylor's expansion. With this simplification, Majidi et al. (2010) could generate an approximate analytical solution. In this work, however, we kept a second nonlinear order in the equations. As a result, it was not possible to generate a full analytical solution, but rather a semi-analytical solution, as detailed in Appendix A.

We found that the accuracy of this simplified solution by Majidi et al. depends on the selected problem parameters. **Fig. 7** shows a comparison between the solutions of Majidi et al. and the proposed solutions for different values of $\alpha$. The flow indexes are, respectively, $n = 0.6$, and 0.4 from left to right. At high values of $\alpha$, both solutions are consistent. However, as $\alpha$ decreases, the approximation error in Majidi's et al. solution increases, as depicted in **Fig. 8**. The error reduces when the flow index increases or approaches unity.

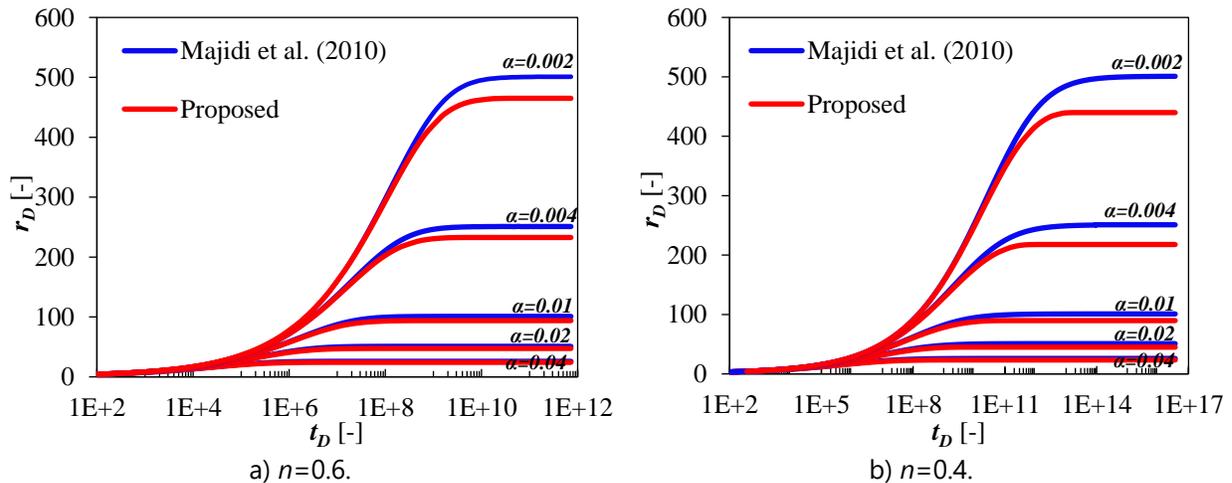

a) *n*=0.6.   b) *n*=0.4.

**Fig. 7**–Comparisons of the dimensionless mud-front radius ($r_D$) versus dimensionless time ($t_D$) obtained by the simplified solutions of Majidi et al. (2010) and our proposed method. Different cases are considered, corresponding to flow indexes *n*=0.60 (a) and *n*=0.40 (b). The simplified solution becomes less accurate as $\alpha$ decreases.

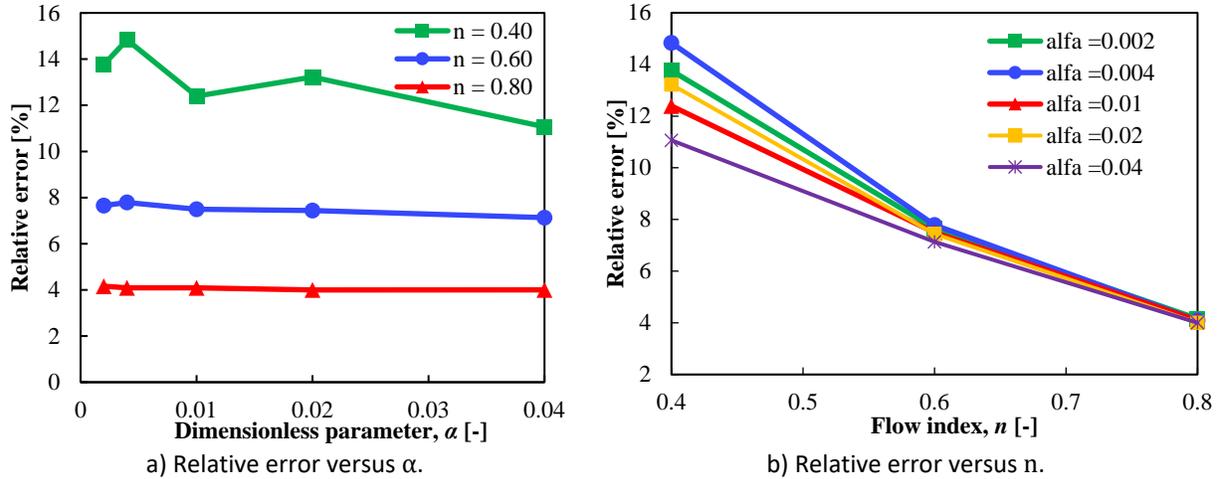

**Fig. 8**–Plots showing the behavior of the deviation error in Majidi et al. (2010) solutions as a function of n (a) and $\alpha$ (b).

## Method Demonstration

The proposed semi-analytical method is applied to four wells from two different fields. In Field case 1, the drilling mud was Bingham plastic fluid with known fluid properties. A deterministic approach is used to match the data with the semi-analytical model. In Field case 2, a more complex Herschel-Bulkley fluid was used. In this case, a probabilistic approach is adopted to account for various uncertainties in the mud and subsurface formation properties. Alterations in the mud properties could occur for various reasons such as mud/water in-situ mixing, transient thermal effects, among others.

### Field Case 1.

In the first case, a drilling Bingham plastic fluid was used for the drilling of two wells: Machar18 and Machar20, in the Machar field in the North Sea (Liétard et al. 2002). Lost circulations within naturally fractured formation were encountered in both wells, as shown in **Fig. 9**. The fracture apertures were estimated to be around 0.42 and 0.64 mm for Machar 18 and Machar 20, respectively, based on a simplified method proposed by (Huang, Griffiths, and Wong 2011). The limitation of Huang's et al. method is in the assumption that the total mud-loss volume must be known, which is available in this case. In our proposed approach, however, the total mud-loss volume can be predicted by fitting the transient mud-loss behavior versus dimensionless time.

We note that the well data, given in Fig. 9, as provided by Liétard et al. 2002, are plotted versus different dimensionless groups than the ones used here. For consistency, we replotted these field data using the proposed dimensionless variables given in Eq. (17), as shown in **Fig. 10**. The semi-analytical solutions were then generated to replicate the transient mud-flow, which produced an excellent match to the trends, as depicted in Fig. 10. In this case, the behavioral flow index is unity (n=1) for Bingham plastic fluid model, 19.5 lb/100ft$^2$ of yield value, and 35 cp of plastic viscosity. The dimensionless parameters are α=0.00215 and 0.0006436, respectively. Fracture apertures of Machar18 and Machar20 wells were found to be $w$ =0.425 mm and 0.616 mm.

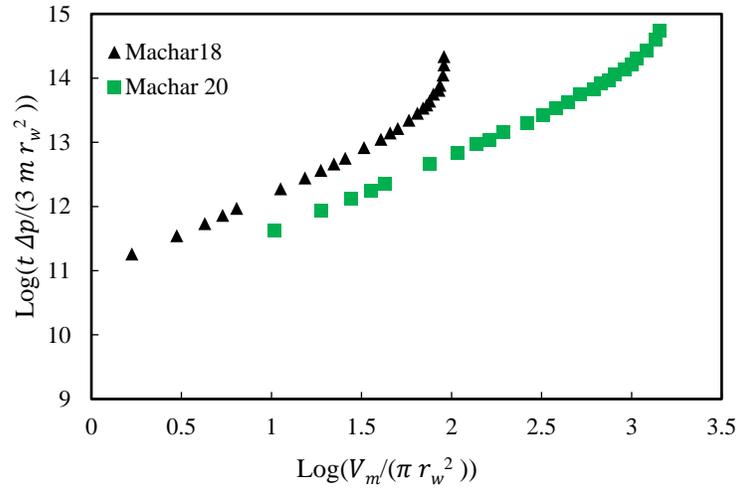

**Fig. 9**–Cumulative mud losses encountered in two wells in the Machar field, data acquired from (Liétard et al. 2002).

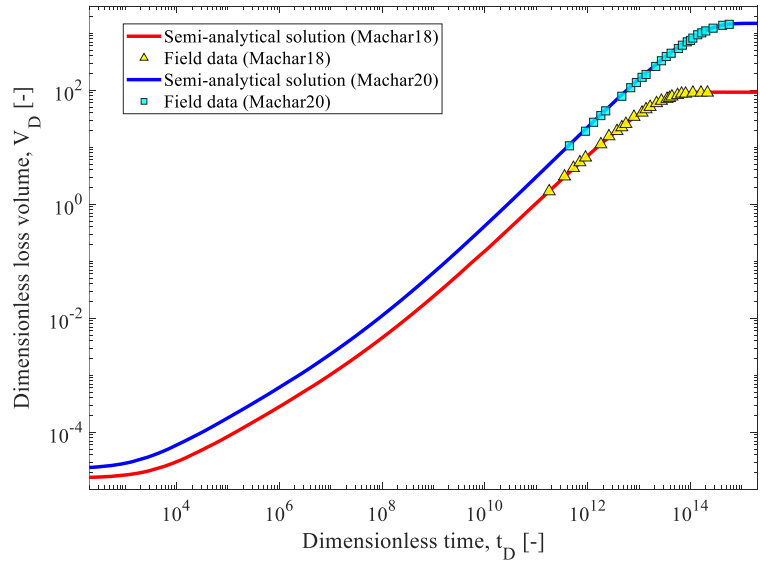

**Fig. 10**–Matching solutions for the two field data sets of Machar field using dimensionless variables. Both graphs data were processed to capture transient leakage behavior. The upper graph is for Machar20, and the lower plot is for Machar18.

**Field Case 2.**

The second field case corresponds to two wells in the Gulf of Mexico (Majidi et al. 2008; Majidl et al. 2010). The mud-loss volumes (gallons per minute) were reported versus time in the two zones for a limited period before the mud-loss stops, as shown in **Fig. 11**. The pressure drop, which is the difference between the injection pressure and initial formation pressure, was reported to be within the range of 700-800 psi, and mud properties are n=0.94, m=0.08 lbf.s/100ft$^2$, and $\tau_0$=8.4 lbf/100ft$^2$ at surface conditions.

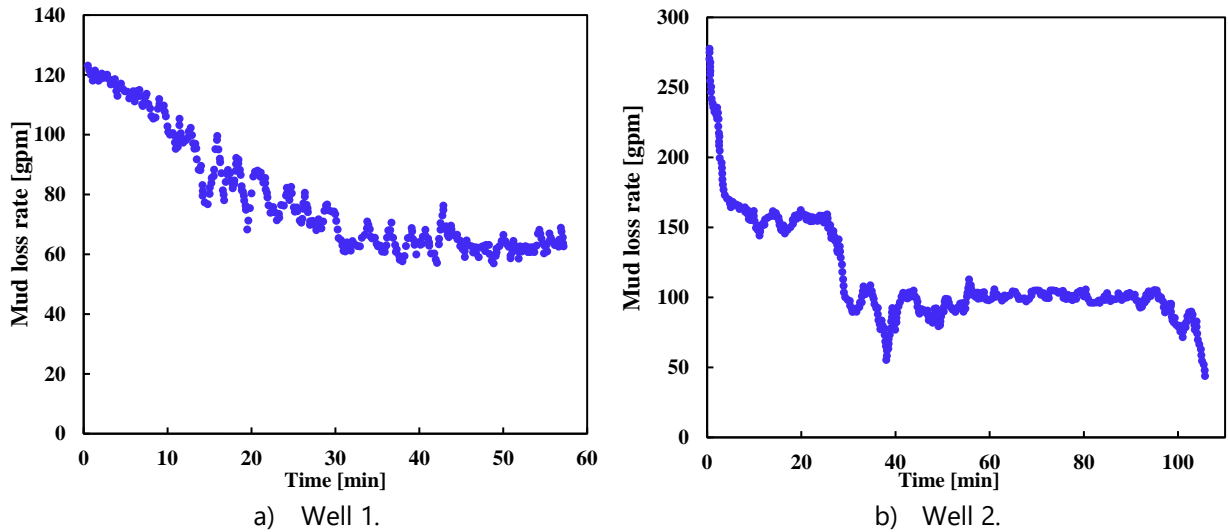

a) Well 1.  
b) Well 2.

**Fig. 11**–Mud-loss volume rate (gallons per minute) versus time (minutes) reported for two wells, 1 and 2, in the Gulf of Mexico (Majidi et al. 2008).

Because of the nature of uncertainties in the drilling fluid and subsurface properties, we apply a probabilistic modeling approach using Monte-Carlo simulations. This probabilistic approach is needed to account for various uncertainties in the fluid property alterations related to subsurface temperature conditions and fluid mixing (Babu 1998; Rommetveit and Bjorkevoll 1997). We perform uncertainty analysis by combining our semi-analytical solutions with Monte-Carlo simulations. The whole process is computationally efficient and could be performed within seconds. We vary six input parameters, including the flow index, yield stress, fracture aperture, consistency factor, and pressure drop.

**Fig. 12** (right) corresponds to the parameters in Zone 2. These distributions are used as inputs for the semi-analytical model, which is driven by the Monte-Carlo simulation process. The history-match corresponds to the time-dependent cumulative mud-loss volume, where uncertainties are taken into consideration. ***Fig. 13*** (left) shows the solution matching band of the field data using the semi-analytical solution combined with Monte-Carlo simulations. The matching p10, p50, and p90 percentiles are also shown. Fig. 13 (right) uses a violin plot to highlight the distribution behavior of the solution versus time. **Fig. 14** shows similar analyses done to match the data in Zone 2.

For Well 1, the obtained probabilistic predictions for the fracture aperture were 0.57, 0.79, and 1.01 *mm* for p10, p50, and p90, respectively. The predicted total mud-loss volumes were 968, 2591, and 5331 bbl for p10, p50, and p90, respectively. The probabilistic distributions of the matching parameters are shown in ***Fig. 15*** for Well 1, and in ***Fig. 16*** for Well 2.

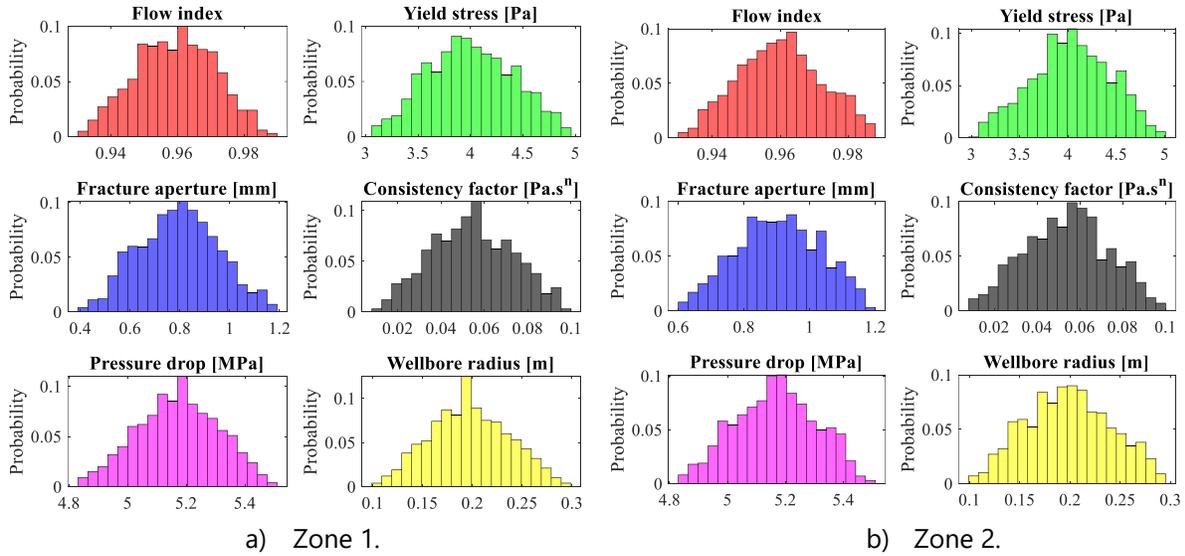

a) Zone 1.  b) Zone 2.

**Fig. 12**–Histograms showing the distributions of the input parameters, modified from (Reza Majidi 2008), used in the Monte-Carlo simulations for Zone 1 (a) and Zone 2 (b).

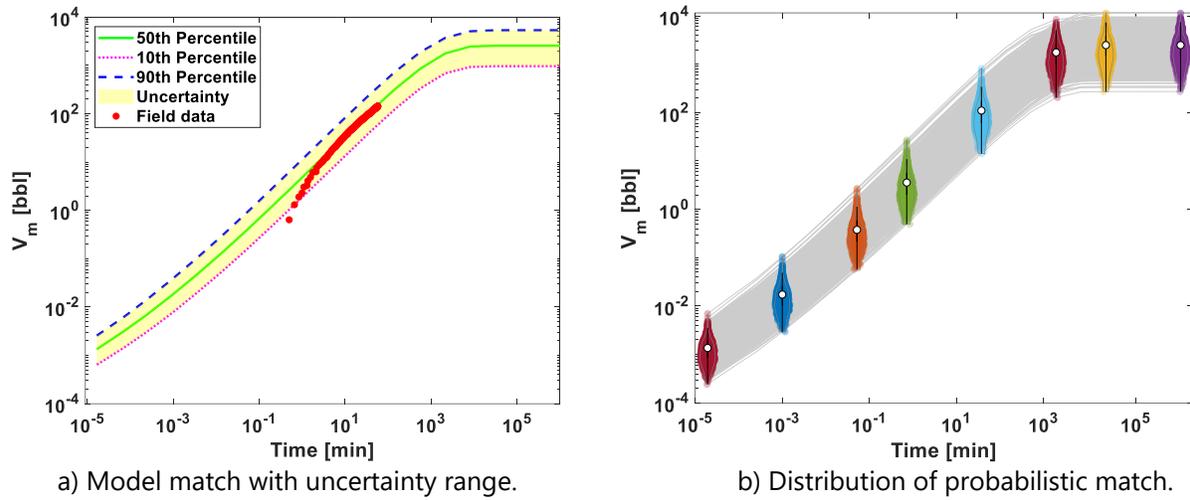

a) Model match with uncertainty range.  b) Distribution of probabilistic match.

**Fig. 13**–Well 1 data and corresponding solution match of cumulative mud-loss volume ($V_m$) versus time, showing the p10, p50, p90, percentiles (a), and with the solution distribution behavior (b).

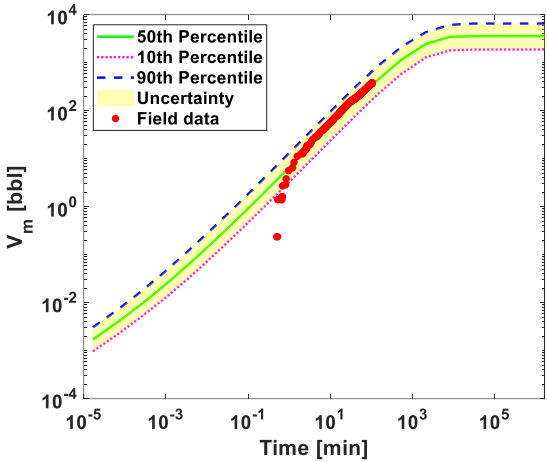
a) Model match with uncertainty range.
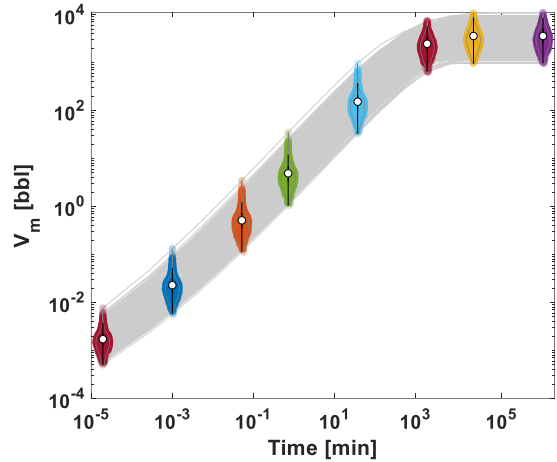
b) Distribution of probabilistic match.

**Fig. 14**–Well 2 data and corresponding solution match of cumulative mud-loss volume versus time, showing the p10, p50, p90, percentiles (a), and with the solution distribution behavior (b).

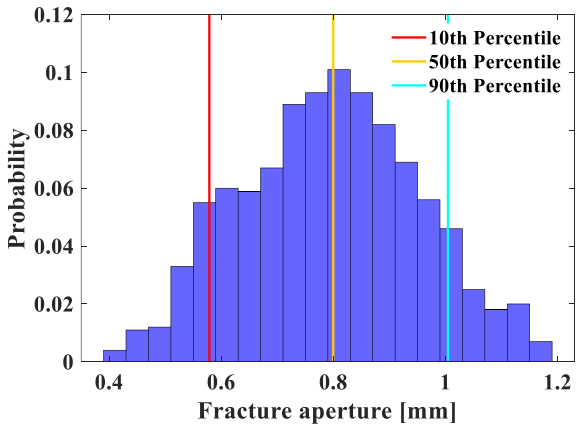
a) Distribution of fracture aperture uncertainty.
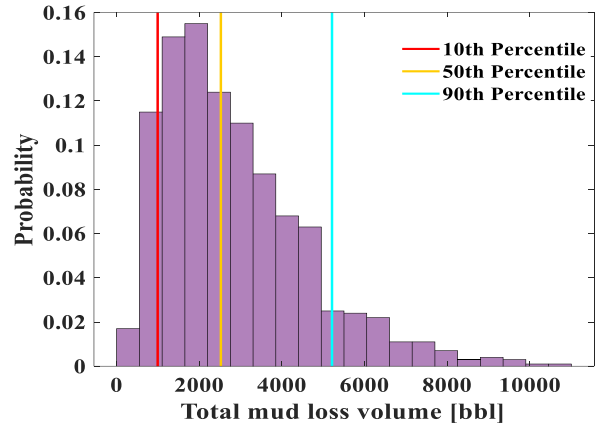
b) Distribution of total mud loss uncertainty.

**Fig. 15**–Histograms showing the predicted probabilistic distribution of the fracture aperture (a) and the total mud-loss volume (b) for Zone 1.

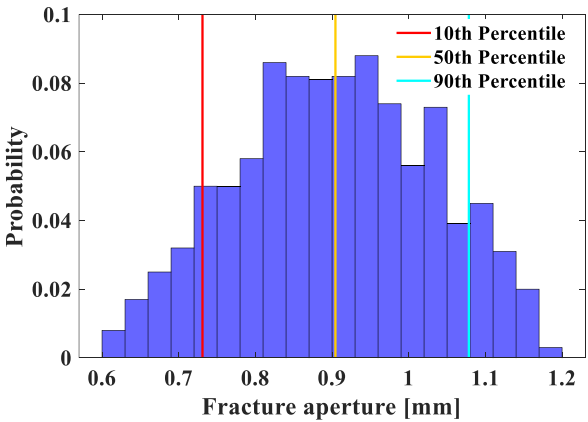
a) Distribution of fracture aperture uncertainty.
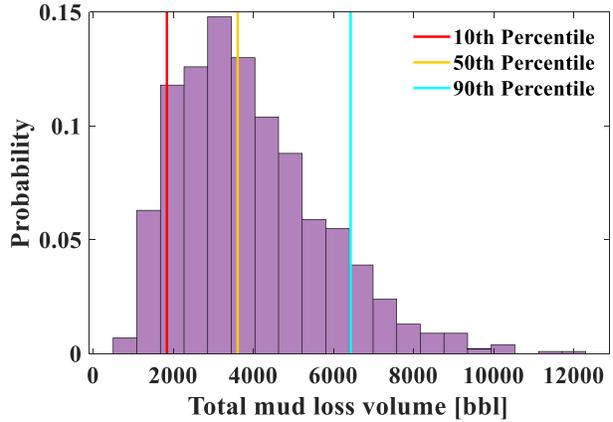
b) Distribution of total mud loss uncertainty.

**Fig. 16**–Histograms showing the predicted probabilistic distribution of the fracture aperture (a) and the total mud-loss volume (b) for Zone 2.

## Discussion

The objective of the proposed model is to develop a numerical diagnostic tool to help predict the dynamic behavior of mud leakage into fractured formation. The tool can be used to perform quick sensitivity analyses and what-if scenarios to optimize LCM selection. The model is based on simple computations and runs efficiently in a spreadsheet. This feature makes it convenient to be coupled with Monte-Carlo simulations to address uncertainties, where thousands of simulations are typically needed. The input data for the model corresponds to the drilling fluid rheological properties, the injection pressure-drop, and the mud-flow rate. The modeling procedure consists of 1) converting and plotting the cumulative mud volumes versus a dimensionless time, and 2) matching the trend with the semi-analytical model or the type-curves by tuning the fracture hydraulic aperture. As a result, the matching trend provides an estimate of the fracture hydraulic aperture and a prediction of the mud-loss behavior, including the flow halt time and the total ultimate volume of mud leakage.

It should be noted that the proposed model is based on several simplifications, and therefore, its applicability is conditional. The main simplifications include the assumption of mud-flow in a horizontal fracture. This assumption was adopted to neglect the buoyancy effect, driven by the difference of densities of the drilling mud and the in-situ water. Consequently, the model is not expected to be applicable when the gravity effect is significant such as in the case of mud with a specific gravity different from one, flowing in inclined fractures. The model also neglects the effect of water displacement, downstream of the mud. Therefore, the pressure of the water phase is assumed constant, equal to the initial pressure of the formation. This assumption is valid when the mud viscosity is significantly higher than the water viscosity. Note that an analogy to this assumption is in Richards equations for water flow in unsaturated porous media, where the air pressure is assumed constant. Other simplifications in our model include ignoring the mud/water in-situ mixing and thermal effects. Furthermore, the modeling procedure may produce nonunique solutions. This problem is common in all curve-fitting approaches, such as pressure transient analysis (PTA) and decline curve analysis (DCA), which mostly occur if the trend of field data is not well established. Therefore, uncertainty analysis should always be accounted for, and different methods should be used to confirm the results.

## Conclusions

Lost circulation during drilling operations is a common problem that requires immediate intervention to circumvent fluid loss. Diagnostic tools, based on simplified input data such as fluid properties, pressure, and rate trends, can be quickly deployed to quantify uncertainties related to the fluid leakage into the subsurface formation and to perform predictions. The main contributions of this work are the following:

- A new semi-analytical approach is presented to model the leakage behavior of general Herschel-Bulkley fluids into a horizontal infinite-acting fracture, mimicking the effect of a fractured formation.
- The analytical approach is based on simplified assumptions such as horizontal fractures with uniform aperture.
- The proposed solution is a generalization of other analytical solutions developed for Bingham plastic fluids. The model is applicable to different types of non-Newtonian fluids, including yield stress shear-thinning and shear-thickening fluids.
- The model was verified for general cases using high-resolution finite element simulations.

- The modeling approach can predict the trend of mud leakage in a system with horizontal fractures as a function of time. It can predict the effective hydraulic aperture of the fracture, the ultimate total mud-loss volume, and the expected duration before the leakage stalls, if conditions allow.
- We introduced dimensionless groups and generated type-curves, which can provide quick diagnostics about the leakage behavior from matching the type-curve trends without a need for simulations.
- We demonstrated the applicability of the model for four wells from two different fields. A numerical procedure was described to couple the model with Monte-Carlo simulations to perform predictions under uncertainties.
- The proposed semi-analytical model is based on simple calculations that can be performed efficiently. This model is useful as a numerical diagnostic tool to perform quick predictions and to help optimize LCM selection by performing what-if scenarios.

## Appendix A: Derivation from Cauchy Equation to transient fluid loss

The Cauchy equation of Motion is given by:

$$\rho \frac{\partial \mathbf{v}}{\partial t} + \rho (\mathbf{v} \cdot \nabla) \mathbf{v} = \nabla \cdot (-p\mathbf{I} + \boldsymbol{\tau}) + \rho \mathbf{g} \quad . \tag{18}$$

At the steady-state without inertial and gravity effects, we get,

$$0 = \nabla \cdot (-p\mathbf{I} + \boldsymbol{\tau}). \tag{19}$$

In a radial system 1D, the above equation simplifies to (see (Panton 1984)):

$$\tau_{rz} = z \frac{\partial p}{\partial r}. \tag{20}$$

Eq.(20) relates the pressure to shear stress. Another equation to relate shear stress to velocity is given by Herschel-Bulkley model, such that,

$$\tau_{rz} = \tau_0 + m \left( \frac{dv_r}{dz} \right)^n \tag{21}$$

Connecting these two Eq.(21) and (20) as,

$$z \frac{\partial p}{\partial r} = \tau_0 + m \left( \frac{dv_r}{dz} \right)^n \tag{22}$$

Solving the differential equation for the general solution of the velocity profile of the half fracture (i.e. $z \in [0, w/2]$) domain, we get,

$$v_r(z) = \frac{n\left(-\frac{\partial p}{\partial r}\frac{w}{2}+\tau_0\right)\left(\frac{\frac{\partial p}{\partial r}\frac{w}{2}-\tau_0}{m}\right)^{1/n} + n\left(-\frac{\partial p}{\partial r}z+\tau_0\right)\left(\frac{\frac{\partial p}{\partial r}z-\tau_0}{m}\right)^{\frac{1}{n}}}{\frac{\partial p}{\partial r}(n+1)} \quad (23)$$

Eq.(23) is valid along the fracture aperture in the z-direction, where two regions, according to the velocity profile, can be distinguished: plug (non-deformed) region and free (deformed) region. We can see from Herschel-Bulkley equation model that the plug region corresponds to zero derivative of the velocity in the z-direction, that is, $\frac{dv_r}{dz} = 0$. Substituting this condition in the fluid model with pressure Eq.(22), the following equation of fluid yield stress is obtained,

$$z_{plug}\frac{\partial p}{\partial r} = \tau_0. \quad (24)$$

The velocity is defined at the three B.C. as,

$$v_r(z) = \begin{cases} v_{r,plug}(z), & \text{for } z \leq z_{plug} \\ v_{r,free}(z), & \text{for } z_{plug} < z < \frac{w}{2} \\ 0, & \text{for } z = \frac{w}{2} \end{cases} \quad (25)$$

In Eq. (5), $z_{plug}$ is the vertical extension of the plug region; $v_{r,plug}$, and $v_{r,free}$ denote the velocities within the plug region and free region, respectively. The last condition is a result of no-slip B.C.

Substitute Eq.(24) into the solution of the general velocity profile in Eq. (23), we get,

$$v_{r,free}(z) = \frac{n}{n+1}\left(z_{plug}-\frac{w}{2}\right)\left(\frac{\frac{\partial p}{\partial r}\left(\frac{w}{2}-z_{plug}\right)}{m}\right)^{1/n} + \frac{n}{n+1}(z-z_{plug})\left(\frac{\frac{\partial p}{\partial r}(z-z_{plug})}{m}\right)^{\frac{1}{n}} \quad (26)$$

Use this Eq.(26) to find velocity profile in the plug region (plug velocity) by substituting $z_{plug} = \frac{\tau_0}{\frac{\partial p}{\partial r}}$ and simplifying, we get,

$$v_{r,plug}(z) = \frac{n}{n+1}\left(\frac{\tau_0}{\frac{\partial p}{\partial r}} - \frac{w}{2}\right)\left(\frac{\left(\frac{w}{2}\frac{\partial p}{\partial r} - \tau_0\right)}{m}\right)^{1/n} \quad (27)$$

Now, we have two velocity profiles as expressed as,

$$v_{r,free}(z) = \frac{n}{n+1}\left(z_{plug} - \frac{w}{2}\right)\left(\frac{\frac{\partial p}{\partial r}\left(\frac{w}{2} - z_{plug}\right)}{m}\right)^{1/n} + \frac{n}{n+1}(z - z_{plug})\left(\frac{\frac{\partial p}{\partial r}(z - z_{plug})}{m}\right)^{\frac{1}{n}}$$

$$v_{r,plug}(z) = \frac{n}{n+1}\left(\frac{\tau_0}{\frac{\partial p}{\partial r}} - \frac{w}{2}\right)\left(\frac{\left(\frac{w}{2}\frac{\partial p}{\partial r} - \tau_0\right)}{m}\right)^{1/n} \quad (28)$$

We transform the velocities into the total volumetric flow rate (flux);

$$Q_{total} = Q_{plug} + Q_{free} \quad (29)$$

Applying the surface integral of the velocity field for the two regions,

$$Q_{total} = 4\pi r \int_0^{z_{plug}} v_{(r)plug} dz + 4\pi r \int_{z_{plug}}^{w/2} v_{(r)free} dz \quad (30)$$

Substituting, we find the total flux,

$$Q_{total} = \frac{4\pi r}{m^{1/n}}\left(\frac{dp}{dr}\right)^{1/n}\left(\frac{w}{2} - \frac{\tau_0}{\frac{dp}{dr}}\right)^{1/n+1}\left(\frac{n}{n+1}\frac{\tau_0}{\frac{dp}{dr}} + \frac{n}{2n+1}\left(\frac{w}{2} - \frac{\tau_0}{\frac{dp}{dr}}\right)\right) \quad (31)$$

Simplifying the above expresses to get,

$$Q_{total} = \frac{4\pi r}{m^{1/n}}\left(\frac{w}{2}\right)^{1/n+2}\left(\frac{n}{2n+1}\right)\left(\frac{dp}{dr}\right)^{1/n}\left(1 - \frac{\tau_0}{\frac{w}{2}\frac{dp}{dr}}\right)^{1/n}\left(1 - \left(\frac{1}{n+1}\right)\frac{\tau_0}{\frac{w}{2}\frac{dp}{dr}} - \left(\frac{n}{n+1}\right)\left(\frac{\tau_0}{\frac{w}{2}\frac{dp}{dr}}\right)^2\right) \quad (32)$$

To avoid obtaining complex number because of the negative $\frac{dp}{dr}$ inside the power term, we rewrite the equation as,

$$Q_{total}^n = \frac{(4\pi r)^n}{m}\left(\frac{w}{2}\right)^{2n+1}\left(\frac{n}{2n+1}\right)^n\left(\frac{dp}{dr}\right)\left(1-\frac{\tau_0}{\frac{w}{2}\frac{dp}{dr}}\right)\left(1-\left(\frac{1}{n+1}\right)\frac{\tau_0}{\frac{w}{2}\frac{dp}{dr}}-\left(\frac{n}{n+1}\right)\left(\frac{\tau_0}{\frac{w}{2}\frac{dp}{dr}}\right)^2\right)^n \quad (33)$$

The last term is approximated by using the second-order Taylor expansion, that is,

$$Q_{total}^n = \frac{(4\pi r)^n}{m}\left(\frac{w}{2}\right)^{2n+1}\left(\frac{n}{2n+1}\right)^n\left(\frac{dp}{dr}\right)\left(1-\frac{\tau_0}{\frac{w}{2}\frac{dp}{dr}}\right)\left(1-\left(\frac{n}{n+1}\right)\frac{\tau_0}{\frac{w}{2}\frac{dp}{dr}}-\left(\frac{n^2}{n+1}\right)\left(\frac{\tau_0}{\frac{w}{2}\frac{dp}{dr}}\right)^2\right) \quad (34)$$

Simplifying,

$$Q_{total}^n = \frac{(4\pi r)^n}{m}\left(\frac{w}{2}\right)^{2n+1}\left(\frac{n}{2n+1}\right)^n\left(\frac{dp}{dr}\right)\left(1-\left(\frac{2n+1}{n+1}\right)\frac{\tau_0}{\frac{w}{2}\frac{dp}{dr}}+\left(\frac{n-n^2}{n+1}\right)\left(\frac{\tau_0}{\frac{w}{2}\frac{dp}{dr}}\right)^2+\frac{n^2}{n+1}\left(\frac{\tau_0}{\frac{w}{2}\frac{dp}{dr}}\right)^3\right)$$

(35)

By neglecting the higher-order terms, we get,

$$Q_{total}^n = \frac{(4\pi r)^n}{m}\left(\frac{w}{2}\right)^{2n+1}\left(\frac{n}{2n+1}\right)^n\left(\frac{dp}{dr}\right)\left(1-\left(\frac{2n+1}{n+1}\right)\frac{\tau_0}{\frac{w}{2}\frac{dp}{dr}}+\left(\frac{n-n^2}{n+1}\right)\left(\frac{\tau_0}{\frac{w}{2}\frac{dp}{dr}}\right)^2\right) \quad (36)$$

We arrange the pressure gradient term as a polynomial equation of second degree, that is,

$$0=\left(\frac{dp}{dr}\right)^2-\left(\frac{Q_{total}^n}{r^n\frac{(4\pi)^n}{m}\left(\frac{w}{2}\right)^{2n+1}\left(\frac{n}{2n+1}\right)^n}+\left(\frac{2n+1}{n+1}\right)\frac{\tau_0}{\frac{w}{2}}\right)\frac{dp}{dr}+\left(\frac{n-n^2}{n+1}\right)\left(\frac{\tau_0}{\frac{w}{2}}\right)^2 \quad (37)$$

This equation can be seen as a quadratic equation with unknown $\frac{dp}{dr}$. To simplify the expression, we define the following quantities,

$$A = \frac{(4\pi)^n}{m} \left(\frac{w}{2}\right)^{2n+1} \left(\frac{n}{2n+1}\right)^n$$

$$B = \left(\frac{2n+1}{n+1}\right) \frac{\tau_0}{w/2} \tag{38}$$

$$D = \left(\frac{n-n^2}{n+1}\right) \left(\frac{\tau_0}{w/2}\right)^2$$

Eq.(37), becomes,

$$0 = \left(\frac{dp}{dr}\right)^2 - \left(\frac{Q_{total}^n}{r^n A} + B\right)\frac{dp}{dr} + D \tag{39}$$

This equation is a nonhomogeneous, nonlinear first-order ordinary differential equation. The general solution of the polynomial quadratic equation is in the form of $\frac{-b \pm \sqrt{b^2 - 4ac}}{2a}$. Hence,

$$\frac{dp}{dr} = \frac{1}{2}\left(B + \frac{Q_{total}^n}{r^n A} \pm \sqrt{\left(B + \frac{Q_{total}^n}{r^n A}\right)^2 - 4D}\right) \tag{40}$$

There are two roots for this equation, from which the positive root is selected, thus,

$$\frac{dp}{dr} = \frac{1}{2}\left(B + \frac{Q_{total}^n}{r^n A} + \sqrt{\left(B + \frac{Q_{total}^n}{r^n A}\right)^2 - 4D}\right) \tag{41}$$

The pressure at the inlet is $p_w$, and the pressure at the mud-font is $p_f$. It should be noted that the interface is moving with time $r_f(t)$. As a result, we implement a moving boundary condition at the interface only, that is,

$$\int_{p_w}^{p_f} dp = \int_{r_w}^{r_f(t)} \frac{1}{2}\left(B + \frac{Q_{total}^n}{r^n A} + \sqrt{\left(B + \frac{Q_{total}^n}{r^n A}\right)^2 - 4D}\right) dr \tag{42}$$

Integrate the left-hand side, and the first two sides on the right-hand side, we get,

$$p_f - p_w = \frac{1}{2}\left(\int_{r_w}^{r_f(t)} B\, dr + \int_{r_w}^{r_f(t)} \frac{Q_{total}^n}{r^n A} dr + \int_{r_w}^{r_f(t)} \left(\sqrt{\left(B + \frac{Q_{total}^n}{r^n A}\right)^2 - 4D}\right) dr\right) \tag{43}$$

$$p_f - p_w = \frac{B(r_f - r_w)}{2} + \frac{Q_{total}^n \left( r_f^{1-n} - r_w^{1-n} \right)}{2(1-n)A} + \frac{1}{2} \int_{r_w}^{r_f(t)} \left( \sqrt{\left( B + \frac{Q_{total}^n}{r^n A} \right)^2 - 4D} \right) dr \qquad (44)$$

To compute the last integral term on the right, we use,

$$\frac{1}{2} \int_{r_w}^{r_f(t)} \left( \sqrt{\left( B + \frac{Q_{total}^n}{r^n A} \right)^2 - 4D} \right) dr = f\left( r_f(t) \right) - f(r_w) \qquad (45)$$

which can be solved as (Choi, Yun, and Choi 2018; Vidunas 2008; Ismail and Pitman 2000),

$$f\left( r_f(t) \right) - f(r_w) =$$

$$\frac{F_1(\alpha, \beta, \beta', \gamma, \chi, \delta) r_f(t) \sqrt{\left( B + \frac{Q_{total}^n}{r_f(t)^n A} \right)^2 - 4D}}{2\sqrt{\dfrac{BQ_{total}^n + Ar_f(t)^n (B^2 - 4D) - 2A\sqrt{\dfrac{Q_{total}^{2n} D}{A^2}}}{r_f(t)^n A(B^2 - 4D)}} \sqrt{\dfrac{BQ_{total}^n + Ar_f(t)^n (B^2 - 4D) + 2A\sqrt{\dfrac{Q_{total}^{2n} D}{A^2}}}{r_f(t)^n A(B^2 - 4D)}}} -$$

$$\frac{F_1(\alpha, \beta, \beta', \gamma, \chi, \delta) r_w \sqrt{\left( B + \frac{Q_{total}^n}{r_w^n A} \right)^2 - 4D}}{2\sqrt{\dfrac{BQ_{total}^n + Ar_w^n (B^2 - 4D) - 2A\sqrt{\dfrac{Q_{total}^{2n} D}{A^2}}}{r_w^n A(B^2 - 4D)}} \sqrt{\dfrac{BQ_{total}^n + Ar_w^n (B^2 - 4D) + 2A\sqrt{\dfrac{Q_{total}^{2n} D}{A^2}}}{r_w^n A(B^2 - 4D)}}} \qquad (46)$$

Appell's hypergeometric function of the first kind is a solution of the integral term. It is a double hypergeometric series $F_1(\alpha, \beta, \beta', \gamma, \chi, \delta)$, where,

$$\alpha = -\frac{1}{n}$$

$$\beta = \beta' = -\frac{1}{2}$$

$$\gamma = \frac{n-1}{n} \qquad (47)$$

$$\chi = -\frac{Q_{total}^n r^{-n}}{AB + 2A\sqrt{D}}$$

$$\delta = -\frac{Q_{total}^n r^{-n}}{AB - 2A\sqrt{D}}$$

We substitute back these values into Eq.(46) to get,

$$f(r_f(t)) - f(r_w) =$$

$$\frac{F_1\left(-\frac{1}{n}, -\frac{1}{2}, -\frac{1}{2}, \frac{n-1}{n}, -\frac{Q_{total}^n r_f(t)^{-n}}{AB + 2\sqrt{D}}, -\frac{Q_{total}^n r_f(t)^{-n}}{AB - 2\sqrt{D}}\right) r_f(t) \sqrt{\left(B + \frac{Q_{total}^n}{r_f(t)^n A}\right)^2 - 4D}}{2\sqrt{\left(1 + \frac{BQ_{total}^n}{r_f(t)^n A(B^2 - 4D)} - \frac{2Q_{total}^n \sqrt{D}}{r_f(t)^n A(B^2 - 4D)}\right)\left(1 + \frac{BQ_{total}^n}{r_f(t)^n A(B^2 - 4D)} + \frac{2Q_{total}^n \sqrt{D}}{r_f(t)^n A(B^2 - 4D)}\right)}} -$$

$$\frac{F_1\left(-\frac{1}{n}, -\frac{1}{2}, -\frac{1}{2}, \frac{n-1}{n}, -\frac{Q_{total}^n r_w^{-n}}{AB + 2\sqrt{D}}, -\frac{Q_{total}^n r_w^{-n}}{AB - 2\sqrt{D}}\right) r_w \sqrt{\left(B + \frac{Q_{total}^n}{r_w^n A}\right)^2 - 4D}}{2\sqrt{\left(1 + \frac{BQ_{total}^n}{r_w^n A(B^2 - 4D)} - \frac{2Q_{total}^n \sqrt{D}}{r_w^n A(B^2 - 4D)}\right)\left(1 + \frac{BQ_{total}^n}{r_w^n A(B^2 - 4D)} + \frac{2Q_{total}^n \sqrt{D}}{r_w^n A(B^2 - 4D)}\right)}}$$

(48)

We now have an analytical solution of pressure as a function of radial distance $r$. The Eq.(48), (45) and (44) are used when the total flow rate entering the fracture is known. Otherwise, if the total flow rate is unknown, we can use the total flow rate as,

$$Q_{total} = \frac{dV_m}{dt} \tag{49}$$

Mud-loss volume can be found for radial flow as,

$$V_m = \pi w \left(r_f(t)^2 - r_w^2\right) \tag{50}$$

Substituting Eq.(50) into (49) and differentiating,

$$Q_{total} = 2\pi w r_f(t) \frac{dr_f(t)}{dt} \tag{51}$$

Finally, the mathematical problem of lost circulation in a smooth horizontal fracture is solved analytically. Eqs.(51), (48), (45) and (44) are the solution of mud-loss flow front $r_f(t)$ as a function of time $t$, based on the given parameters of mud rheology ($n$, $m$, $\tau_0$), differential pressure ($p_f$, $p_w$) and fracture aperture ($w$).

## Appendix B: Particular solution for Bingham plastic fluids

Here, we show how the proposed solution can be derived mathematically for Bingham plastic fluids, when $n=1$. Let us take Eq.(36), after approximating the last term, by applying the same approach for Bingham plastic fluid condition ($n=1$),

$$Q_{total} = \frac{\pi r w^3}{6m}\left(\frac{dp}{dr} - \frac{3\tau_0}{w}\right) \tag{52}$$

Integrating,

$$\int_{p_w}^{p_f} dp = \int_{r_w}^{r_f(t)} \left( \frac{6mQ_{total}}{\pi r w^3} + \frac{3\tau_0}{w} \right) dr \tag{53}$$

The above equation becomes,

$$p_f - p_w = \frac{12 m r_f(t)}{w^2} \frac{dr_f(t)}{dt} \ln\left(\frac{r_f(t)}{r_w}\right) + \frac{3\tau_0}{w}\left(r_f(t) - r_w\right) \tag{54}$$

Rearranging,

$$\frac{dr_f(t)}{dt} = \frac{(p_f - p_w) - \frac{3\tau_0}{w}(r_f(t) - r_w)}{\frac{12 m r_f(t)}{w^2} \ln\left(\frac{r_f(t)}{r_w}\right)} \tag{55}$$

Eq.(55) is the final solution of Bingham plastic fluid model of mud invasion front $r_f(t)$ as a function of time $t$. The equation can be derived in different forms, which is consistent with the dimensionless form by (Liétard et al. 2002). The defined dimensionless groups are,

$$\begin{aligned}
\alpha &= \frac{3 r_w}{w} \left( \frac{\tau_0}{p_f - p_w} \right) \\
\beta &= \left(\frac{w}{r_w}\right)^2 \frac{(p_f - p_w)}{12m} \\
r_D &= \frac{r_f}{r_w}, \quad t_D = \beta t
\end{aligned} \tag{56}$$

Implementing into Eq.(55) to generate a dimensionless differential form,

$$\frac{dr_D(t)}{dt_D} = \frac{1 - \alpha(r_D(t) - 1)}{r_D(t) \ln(r_D(t))} \tag{57}$$

This is an ordinary differential equation that can be solved analytically by using Inverse Function Theorem (IFT) in such a way that any function $f(x)$ is both differentiable and invertible. Assuming $y = f^{-1}(x)$ is the inverse of $f(x)$ which all $x$ satisfying $f'(f^{-1}(x)) \neq 0$. Thus, the derivative,

$$\frac{dy}{dx} = \frac{d}{dx}\left(f^{-1}(x)\right) = \left(f^{-1}\right)'(x) = \frac{1}{f'(f^{-1}(x))} \tag{58}$$

Eq.(57) becomes,

$$\frac{dt_D}{dr_D(t)} = \frac{r_D(t)\ln(r_D(t))}{1-\alpha(r_D(t)-1)} \tag{59}$$

Providing the closed-form solution after applying initial condition at $t_D = 0$ and $r_D = 1$, we get,

$$t_D = -\frac{4}{\alpha}\left(1-r_D+\ln\left(r_D^{r_D}\right)\right)+4\left(\frac{1}{\alpha^2}+\frac{1}{\alpha}\right)\left[\mathbf{Li}_2\left(\frac{\alpha}{1+\alpha}\right)-\mathbf{Li}_2\left(\frac{\alpha r_D}{1+\alpha}\right)-\ln\left(r_{Df}\right)\ln\left(1-\frac{\alpha r_D}{1+\alpha}\right)\right] \tag{60}$$

In the above equation, the polylogarithmic function $\mathbf{Li}_2$ is shown as a solution of Eq.(59), which is the analytical solution of a dimensionless mud invasion front $r_D$ as a function of dimensionless time $t_D$ for Bingham Plastic fluid model. Furthermore, an analytical solution was provided as well (Liétard et al. 2002, 2002).

## Appendix C: Type-cures

We derive the type-Curves in dimensionless form. Let us recall Eq.(44)

$$p_f - p_w = \frac{B(r_f - r_w)}{2} + \frac{Q_{total}^n\left(r_f^{1-n}-r_w^{1-n}\right)}{2(1-n)A} + \frac{1}{2}\int_{r_w}^{r_f(t)}\left(\sqrt{\left(B+\frac{Q_{total}^n}{r^n A}\right)^2 - 4D}\right)dr$$

Dividing all terms by $\Delta p = p_f - p_w$, we get,

$$1 = \frac{B(r_f - r_w)}{2\Delta p} + \frac{Q_{total}^n\left(r_f^{1-n}-r_w^{1-n}\right)}{2\Delta p(1-n)A} + \frac{1}{2\Delta p}\int_{r_w}^{r_f(t)}\left(\sqrt{\left(B+\frac{Q_{total}^n}{r^n A}\right)^2 - 4D}\right)dr \tag{61}$$

Substituting the defined constants as done previously in Eq.(38) and applying the dimensionless variable $r_D = \frac{r_f}{r_w}$, it yields,

$$1 = \frac{1}{2}\left(\frac{2n+1}{n+1}\right)\left(\frac{\tau}{\Delta p}\right)\left(\frac{2r_w}{w}\right)(r_D-1) + 2^n\left(r_D^{1-n}-1\right)\left(\left(\frac{2n+1}{n}\right)\left(\frac{m}{\Delta p}\right)^{\frac{1}{n}}\left(\frac{r_w}{w}\right)^{\frac{n+1}{n}} r_D r_D'\right)^n \tag{62}$$

The transforming of the third term to dimensionless form, we get,

$$\frac{1}{2}\int_{r_w}^{r_f(t)}\left(\sqrt{\left(\frac{B}{\Delta p}+\frac{Q_{total}^n}{\Delta p r^n A}\right)^2 - \frac{4D}{\Delta p^2}}\right)dr = \tag{63}$$

$$\frac{F_1\left(-\frac{1}{n},-\frac{1}{2},-\frac{1}{2},\frac{n-1}{n},-\frac{\left(\beta\frac{dr_D}{dt}\right)^n}{1+2\sqrt{\frac{n-n^2}{n+1}}+2n\left(1+\sqrt{\frac{n-n^2}{n+1}}\right)},-\frac{\left(\beta\frac{dr_D}{dt}\right)^n}{2\sqrt{\frac{n-n^2}{n+1}}-1+2n\left(\sqrt{\frac{n-n^2}{n+1}}-1\right)}\right)\alpha r_D\sqrt{4n(n^2-1)+\left(1+2n+\left(\beta\frac{dr_D}{dt}\right)^n\right)^2}}{2\sqrt{\left(1+\frac{2(2n+1)}{(1+4n^2+4n^3)}\left(\beta\frac{dr_D}{dt}\right)^n+\frac{4n(n^2-1)}{(1+3n+4n^2+n^3)^2}\left(\beta\frac{dr_D}{dt}\right)^{2n}+\frac{(2n+1)^2}{(1+4n^2+4n^3)^2}\left(\beta\frac{dr_D}{dt}\right)^{2n}\right)}}-$$

$$\frac{F_1\left(-\frac{1}{n},-\frac{1}{2},-\frac{1}{2},\frac{n-1}{n},-\frac{\left(\beta r_D\frac{dr_D}{dt}\right)^n}{1+2\sqrt{\frac{n-n^2}{n+1}}+2n\left(1+\sqrt{\frac{n-n^2}{n+1}}\right)},-\frac{\left(\beta r_D\frac{dr_D}{dt}\right)^n}{2\sqrt{\frac{n-n^2}{n+1}}-1+2n\left(\sqrt{\frac{n-n^2}{n+1}}-1\right)}\right)\alpha_D\sqrt{4n(n^2-1)+\left(1+2n+\left(\beta_D r_D\frac{dr_D}{dt}\right)^n\right)^2}}{2\sqrt{\left(1+\frac{2(2n+1)}{(1+4n^2+4n^3)}\left(\beta r_D\frac{dr_D}{dt}\right)^n+\frac{4n(n^2-1)}{(1+3n+4n^2+n^3)^2}\left(\beta r_D\frac{dr_D}{dt}\right)^{2n}+\frac{(2n+1)^2}{(1+4n^2+4n^3)^2}\left(\beta r_D\frac{dr_D}{dt}\right)^{2n}\right)}}$$

Hence, the proposed dimensionless parameters are,

$$r_D = \frac{r_f}{r_w}$$

$$t_D = t\beta$$

$$V_D = \frac{V_m}{V_w} = \frac{\pi w(r_f^2 - r_w^2)}{\pi w r_w^2} = \left(\frac{r_f}{r_w}\right)^2 - 1 = r_D^2 - 1 \qquad (64)$$

$$\alpha = \left(\frac{2n+1}{n+1}\right)\left(\frac{2r_w}{w}\right)\left(\frac{\tau_0}{\Delta p}\right)$$

$$\beta = \left(\frac{n}{2n+1}\right)\left(\frac{w}{r_w}\right)^{1+\frac{1}{n}}\left(\frac{\Delta p}{m}\right)^{\frac{1}{n}}$$

## Nomenclature

| | | |
|---|---|---|
| **v** | = | velocity vector, m/s |
| $A$ | = | defined constant |
| $B$ | = | defined constant |
| $D$ | = | defined constant |
| $d$ | = | normal derivative |
| $O$ | = | truncation error |
| $F_1(\alpha,\beta,\beta',\gamma,\chi,\delta)$ | = | appel's series |
| **g** | = | gravitational acceleration, m/s² |
| **I** | = | identity matrix |
| **Li**$_2$ | = | polylogarithmic function |
| $m$ | = | consistency multiplier, kg/(Pa.s$^n$) |
| $m_p$ | = | regularization exponent, s |
| $n$ | = | flow behavioral index |
| $p$ | = | pressure, psi |
| $p_f$ | = | formation pressure, psi |
| $p_w$ | = | wellbore pressure, psi |
| $Q_{free}$ | = | volumetric flow rate in free region, m³/s |
| $Q_{plug}$ | = | volumetric flow rate in plug region, m³/s |
| $Q_{total}$ | = | total volumetric flow rate, m³/s |
| $r$ | = | radial distance, m |
| $r_D$ | = | dimensionless mud front radius, m |
| $r_f$ | = | mud front radius, m |
| $r_w$ | = | wellbore radius, m |
| $t$ | = | time, min |
| $t_D$ | = | dimensionless time |
| $V_D$ | = | dimensionless mud loss volume |
| $V_m$ | = | cumulative mud loss volume, bbl |
| $V_w$ | = | wellbore volume due to fracture aperture, bbl |
| $v_r$ | = | radial velocity, m/s |
| $v_{r,plug}$ | = | radial velocity in plug region (non-deformed region), m/s |
| $v_{r,free}$ | = | radial velocity profile in the free deformed region, m/s |
| $w$ | = | fracture aperture, mm |
| $z$ | = | z-direction in radial coordinate, m |
| $z_{plug}$ | = | height of plug region profile, m |
| $\alpha$ | = | dimensionless parameter |
| $\beta$ | = | dimensionless parameter |
| $\partial$ | = | partial derivative |
| $\gamma$ | = | shear rate, 1/s |
| $\Delta p$ | = | pressure drop, psi |
| $\mu_0$ | = | viscosity due to fluid yield stress, Pa.s |
| $\mu_{eff}$ | = | effective viscosity, Pa.s |
| $\rho$ | = | density |
| $\boldsymbol{\tau}$ | = | shear stress tensor, Pa |
| $\tau(z,r)$ | = | shear stress component as a function of z and r, Pa |
| $\tau$ | = | shear stress, Pa |
| $\tau_0$ | = | fluid yield stress, Pa |

## Acknowledgments

The authors thank King Abdullah University of Science and Technology (KAUST), and Ali I. Al-Naimi Petroleum Engineering Research Center (ANPERC) for supporting this work.